\documentclass[]{aa} % for a paper on 1 column  

\usepackage{graphicx}
\usepackage{txfonts}
\begin{document}

   \title{A non-glitch speed-up event in the Crab Pulsar}

   \author{M. Vivekanand}

   \institute{No. $24$, NTI Layout $1$\textsuperscript{st} Stage, 
              $3$\textsuperscript{rd} Main, $1$\textsuperscript{st} 
              Cross, Nagasettyhalli, Bangalore $560094$, India. \\
              \email{viv.maddali@gmail.com}
             }

   \date{Received ---; accepted ---}

% \abstract{}{}{}{}{} 
% 5 {} token are mandatory
 
  \abstract
   { 
     The rotation history of the Crab Pulsar is well described by (1) a rotation frequency 
     $\nu$ and a slowdown model that is specified by its first two time derivatives $\dot 
     \nu$ and $\ddot \nu$, known as the secular slowdown model, (2) occasional (once in 
     $\approx 2$ years) significant and abrupt increases in the magnitude of $\nu$ and $\dot 
     \nu$ (occurring on timescales of minutes), known as glitches, and (3) much slower 
     increases and decreases in $\nu$ and $\dot \nu$ (occurring over months and years)
     that are an order  of magnitude smaller, known as timing noise.
   }
   {
     This work reports a speed-up event in the Crab Pulsar that occurred around $2015$ February 
     that is distinct from glitches and timing noise.
   }
   {
     Monthly $\nu$s and $\dot \nu$s of the Crab Pulsar, obtained at radio frequencies and published 
     by Jodrell Bank Observatory (JBO), are used to demonstrate the speed-up event. Monthly 
     arrival times of the  Crab Pulsar's pulse, also published by JBO, combined with X-ray data 
     from the RXTE, SWIFT, and NUSTAR observatories are used to verify the result.
   } % results heading (mandatory)
   {
     The speed-up event is caused by a persistent increase in $\dot \nu$, which results in 
     a monotonic increase in $\nu$. Over the last $\approx 550$ days, $\nu$ has increased 
     monotonically by an amount that is $\approx 10$ times larger than the timing noise level.
   }
   {
    This is a unique event in the Crab Pulsar. This is probably due to a small increase in the Crab 
    Pulsar's internal temperature. In its absence, the next large glitch in the Crab Pulsar is
    expected to occur around $2019$ March. However, this event could have an important bearing 
    on its occurrence.
   }


   \maketitle
%
%________________________________________________________________

\section{Introduction}

Recently \cite{Lyne2015} discussed the rotation history of the Crab Pulsar over the last $45$
years. By studying the three best (isolated and large) glitches among the $24$ that have 
been observed in the Crab Pulsar so far, they show that the apparently abrupt decrease in 
$\dot \nu$ at a glitch (increase in magnitude of negative value) actually has a detail:
only about half the decrease occurs instantaneously; the rest occurs asymptotically 
quasi exponentially on a timescale of $\approx 320$ days; see Figure $3$ of 
\cite{Lyne2015}. The three best glitches were chosen by the criteria that a change in 
the magnitude of $\nu$ and $\dot \nu$ at the glitch should be large, and also by the criteria 
that  the previous and 
subsequent glitches should occur at least $800$ days before and $1200$ days after each 
glitch, respectively.  However, for the glitch of 2011 November 
(at MJD $55875.5$; hereafter CPG2011), they only had data  for $\approx 
800$ days after the glitch, This work analyzes the additional data that has since been 
published by JBO, which reveals a phenomenon unreported so far in Crab or any other 
Pulsar.

\begin{figure}[h]
\centering
\includegraphics[width=8.5cm]{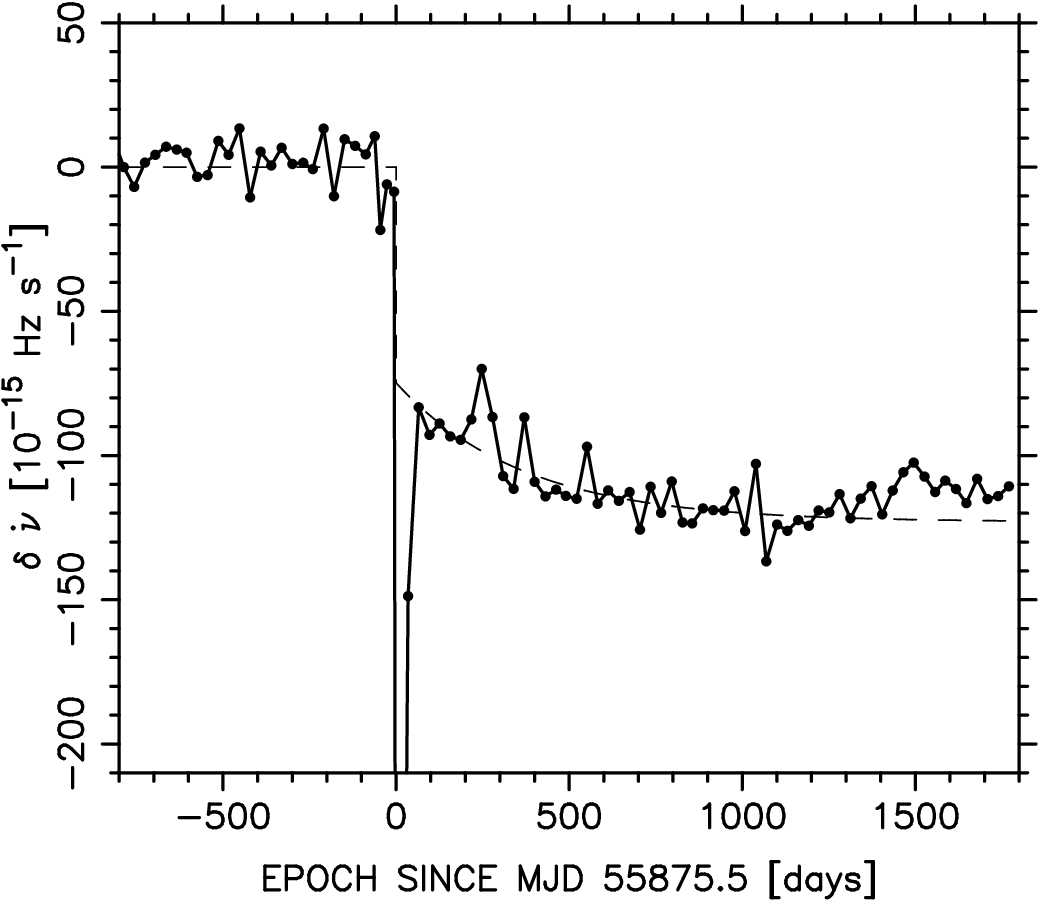}
\caption{
         Frequency derivative residual $\delta \dot \nu$ plotted against epoch 
         since CPG2011. The dashed line before CPG2011 represents the straight line
         fit to the $\dot \nu$ data at negative abscissa. The dashed curve after 
         CPG2011 represents the best fit of the model of \cite{Lyne2015}.
        }
\label{fig1}
\end{figure}

Figure~\ref{fig1} is similar to panel $3$ of Figure $3$ in \cite{Lyne2015}. It is obtained
from the frequency derivative $\dot \nu$ values tabulated in the so-called Jodrell Bank 
Crab Pulsar Monthly 
Ephemeris\footnote{http://www.jb.man.ac.uk/pulsar/crab.html} (\cite{Lyne1993}; hereafter 
JBCPME).  This paper focuses on the significant departure of the data from the model curve in 
Figure~\ref{fig1}, starting $\approx 1200$ days after CPG2011  and lasting until now 
($2016$ September $15$). This implies a persistent and systematic increase in $\delta \dot \nu$ 
with respect to the model of \cite{Lyne2015}, which they consider to be the prototypical 
behavior of all glitches in the Crab Pulsar.

Figure~\ref{fig1} is obtained by fitting a straight line to the $28$ $\dot \nu$ from 
JBCPME, as a function of epoch, for the $800$ days  before CPG2011, resulting in 
a $\dot \nu_0$ value of $-370730(2) \times 10^{-15}$ Hz s$^{-1}$ at the glitch epoch; 
the error in the last digit is shown in brackets. The slope $\ddot \nu_0$ is $1.182(6) 
\times 10^{-20}$ Hz s$^{-2}$. Subtracting the straight line from the $\dot \nu$ values 
results in the $\delta \dot \nu$ shown as dots in Figure~\ref{fig1}. The well-studied 
glitch behavior of the Crab Pulsar implies that the $\delta \dot \nu$ data from days $0$ 
to $1200$ should ideally be fit to the model
\begin{equation}
\delta \dot \nu(t)  = - \Delta \dot \nu_n \exp-(t/\tau_1) + 
                 \Delta \dot \nu_p \left ( w \exp-(t/\tau_2) - 1 \right ),
\end{equation}
\noindent the second term representing the longtime recovery proposed by \cite{Lyne2015} 
in their equation $6$. Now the short recovery timescale $\tau_1$ is typically $\approx 
10$ days. However, the JBCPME has only one $\dot \nu$ value within $13$ days of 
CPG2011, and only two values within $34$ days of CPG2011, so the cadence of data is 
too poor to fit to the first term in Eq. $1$. Furthermore, the errors on these two 
$\dot \nu$ are a factor of $\approx 10$ to $30$ larger than on the rest of the data. 
So the data from $100$ to $1200$ days was fit only to the second term in equation $1$.  
The results are $\Delta \dot \nu_p = 129(6) \times 10^{-15}$ Hz s$^{-1}$, $w = 0.39(3)$ 
and $\tau_2 = 510 \pm 197$ days. By fitting up to $1050$ days only, one obtains $\Delta 
\dot \nu_p = 123(6) \times 10^{-15}$ Hz s$^{-1}$, $w = 0.39(5)$, and $\tau_2 = 367 \pm 
172$ days. These values are consistent with those derived by \cite{Lyne2015}, which are 
$\Delta \dot \nu_p = 132(5) \times 10^{-15}$ Hz s$^{-1}$, $w = 0.46$, and $\tau_2 = 320 
\pm 20$ days, although the errors on $\tau_2$ are very large. In both cases the 
persistent increase in $\delta \dot \nu$ starting $\approx 1200$ days after CPG2011, of
$\approx 11(1) \times 10^{-15}$ Hz s$^{-1}$, is very evident. The dashed curve in 
Figure~\ref{fig1} at positive abscissa is obtained using the latter set of parameters.

A persistent increase in $\delta \dot \nu$ should result in a monotonically increasing
frequency residual $\delta \nu$, which is the integral of $\delta \dot \nu$. This is 
demonstrated in the next section.

\section{Analysis of $\nu$}

\begin{figure}[b]
\centering
\includegraphics[width=8.5cm]{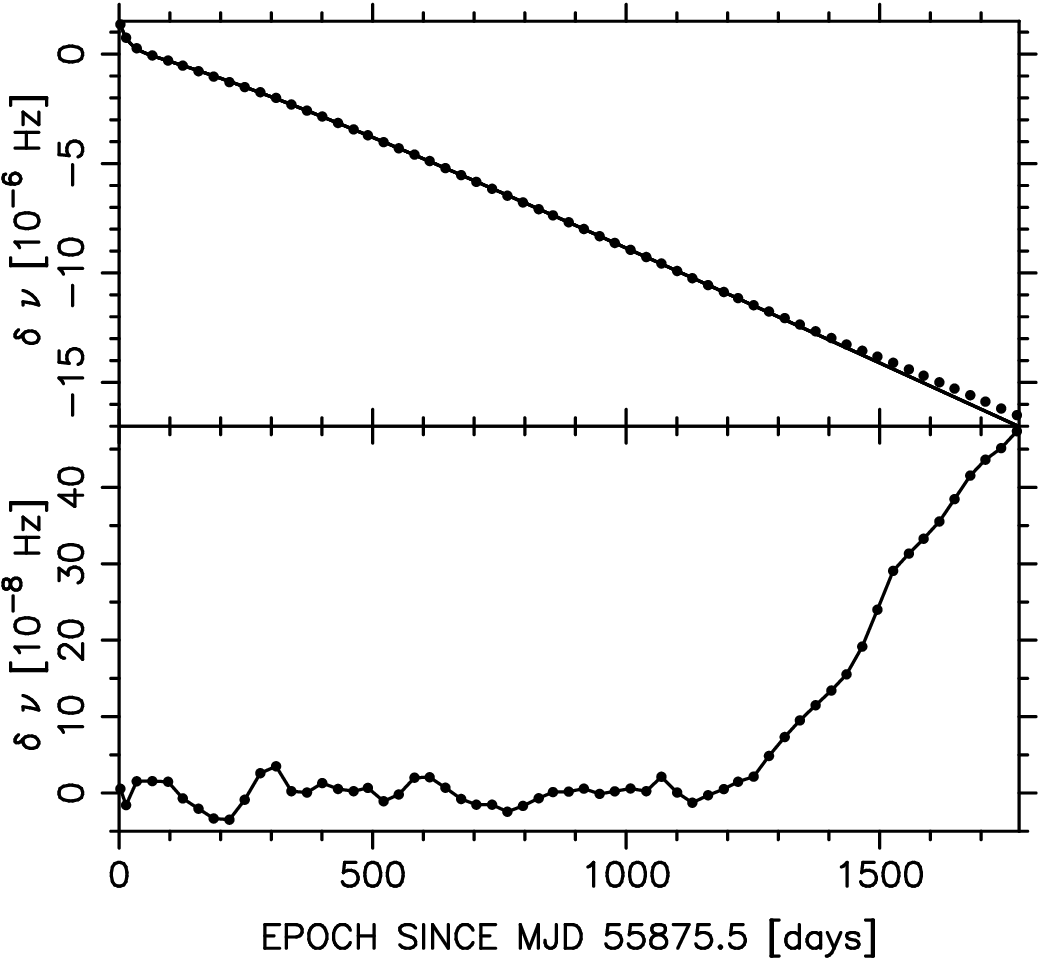}
\caption{
         Top panel: Frequency residuals $\delta \nu$ plotted against epoch since CPG2011.  
         The curve represents the best fit model given in equation $2$. Bottom panel: 
         Difference between the data and the model curve in the top panel. The positive 
         departure of data from the model beyond $\approx 1200$ days is now clearly visible. 
        }
\label{fig2}
\end{figure}

The $28$ $\nu$ values from JBCPME, for the $800$ days  before CPG2011, were fit to a 
quadratic curve as a function of epoch. The results are $\nu_0 = 29.706643782(8)$ Hz at 
the glitch epoch, the first and second derivatives being $\dot \nu_0 = -370727.7(5) \times 
10^{-15}$ Hz s$^{-1}$ and $\ddot \nu_0 = 1.179(1) \times 10^{-20}$ Hz s$^{-2}$. The last 
two parameters are statistically consistent with those derived in Sect. $1$. Subtracting 
this quadratic model from the $\nu$ data results in the $\delta \nu$ shown as dots in 
the top panel of Fig.~\ref{fig2}. The data from $0$ to $1200$ days were fit to the 
model
\begin{equation}
\begin{array}{ll}
\delta \nu(t)  & = \Delta \nu_p + \Delta \nu_n \exp-(t/\tau_1) \\
               & - \Delta \dot \nu_p \left ( w \tau_2 \left ( \exp-(t/\tau_2) - 
                 1 \right ) + t \right ), \\
\end{array}
\end{equation}
\noindent which is an integral of equation $1$ with some terms redefined. In both equations
the subscripts $_p$ and $_n$ refer to permanent and exponentially decaying changes,
respectively,  in the corresponding parameters; see \cite{Shemar1996} and \cite{Vivekanand2015} 
for details. The results for three of the parameters are $\Delta \dot \nu_p = 123(1) \times 
10^{-15}$ Hz s$^{-1}$, $w = 0.44(2)$, and $\tau_2 = 317 \pm 25$ days, which are consistent
with the values derived in section $1$, and  with the values of \cite{Lyne2015}. The
other three parameters are $\Delta \nu_p = 3.2(2) \times 10^{-7}$ Hz, $\Delta \nu_n = 12.2(3)
\times 10^{-7}$ Hz, and $\tau_1 = 16 \pm 1$ days. The step change in $\nu$ at CPG2011 is
$\left ( \Delta \nu_p + \Delta \nu_n \right ) \times 10^{+7} = 15.4(4)$ Hz, which compares 
well with the value of $14.6(1)$ obtained by \cite{Lyne2015}. The step change in $\dot \nu$ at 
CPG2011 is $\left (- \Delta \nu_n / \tau_1 - \Delta \dot \nu_p \times ( -w + 1) \right ) \times
10^{+15} = -951 \pm 59$ Hz $s^{-1}$. This number has not been given by \cite{Lyne2015}.

Although equation $2$ is the integration of equation $1$, two parameters of the latter
could only be derived using equation $2$ for reasons of cadence and large errors. Furthermore,
the parameter $\Delta \nu_p$ is the integration constant that does not exist in equation $1$.

In the top panel of Figure~\ref{fig2} the model departs from the data from epoch $\approx 
1200$ days onwards. This stands out strongly in the bottom panel, which displays 
the difference between the data and the model. After secular slowdown and glitches have 
been accounted for in the timing behavior of the Crab Pulsar, one expects to see only timing 
noise, which is evident before $\approx 1200$ days in the bottom panel. Here $\delta \nu$ 
varies on timescales of $\approx 100$ days with an rms magnitude of $\approx 1.5 \times 
10^{-8}$ Hz. However, the speed-up event causes $\delta \nu$ to increase monotonically to 
$47.4 \times 10^{-8}$ Hz in $\approx 550$ days. Clearly a monotonic variation that is 
$\approx 30$ times larger than the rms cannot be due to timing noise. Given the typical 
monthly cadence of JBCPME data, one can specify the exact epoch of occurrence, and the duration,
of this speed-up event only to an accuracy of one month.

Monotonically increasing frequency residuals $\delta \nu$ should result in monotonically
decreasing residuals of pulse phase, since an increase in frequency leads to a decrease 
in phase in the TEMPO2 package (see \citealt{Shemar1996, Vivekanand2015}). This is discussed 
in the following two sections.

\section{Observations of times of arrival}

Times of arrival (TOA) of the main peak of the Crab Pulsar are also tabulated in the JBCPME, 
referred to the solar system barycenter, and scaled to infinite frequency. Eighty-eight of these 
TOA were combined with TOA from the following three X-ray observatories.

\subsection{RXTE Observatory}

Fifty-seven observation identification numbers (ObsID) are used from the Proportional Counter 
Array (PCA; \citealt{Jahoda1996}) of RXTE, the first obtained on $2009$ September $12$ (ObsID 
$94802$-$01$-$16$-$00$), and the last on $2011$ December $31$ (ObsID $96802$-$01$-$21$-$00$). 
The data (with event mode identifier \textit{E\_250us\_128M\_0\_1s}) and their analysis 
are described in detail in \cite{Vivekanand2015, Vivekanand2016a, Vivekanand2016b}.

\subsection{SWIFT Observatory}

Forty-four ObsID from the X-Ray Telescope (XRT; \citealt{Burrows2005}) on board the SWIFT observatory 
(\citealt{Gehrels2004}) were analyzed; data were obtained in the \textit{wt} mode, which has 
a time resolution of $1.7791$ milliseconds (ms). The first observation was obtained 
on $2009$ September $17$ (ObsID $00058990010$) and the last on $2016$ April $01$ (ObsID 
$00080359006$). The \textit{TIMEPIXR} keyword was set to the value $0.5$ (see XRT 
digest\footnote{http://www.swift.ac.uk/analysis/xrt/digest\_sci.php}). The tool 
\textit{xrtpipeline} was run with the coordinates of the Crab Pulsar. The rest of the analysis 
was as described in \cite{Vivekanand2015}. The tool \textit{barycorr} was used for barycentric 
correction. Pile up in general is not an issue for pulsar timing, since its main effect
is to distort the spectrum, and not to affect the arrival times of photons. 
%\LEt{ Pile up  is generally not an issue for pulsar timing, since its main effect
%is to distort the spectrum and not to determine (?) the arrival times of photons.  }

\subsection{NUSTAR Observatory}

Thirty-eight ObsID from the NUSTAR observatory (\citealt{Harrison2013}) were analyzed; they  had 
live times of at least $\approx 1000$ seconds. The first observation was obtained on
$2012$ September $20$ (ObsID $10013021002$) and the last on $2014$ October $02$ (ObsID 
$10002001008$). The tools \textit{nupipeline} and \textit{barycorr} were used. The 
dead time corrected pulse profile was obtained by using the live time data in the
\textit{PRIOR} column (\citealt{Madsen2015}). The rest of the analysis was as described 
in \cite{Vivekanand2015}.

\section{Analysis of times of arrival}

The combined $227$ TOA in a duration of $\approx 1770$ days yields a mean cadence of once in
$\approx 8$ days, which is a significant improvement over that of JBCPME. All data 
have been barycenter corrected using the same ephemeris (DE200). The published phase 
offsets between the X-ray and radio pulses were inserted for the RXTE (\citealt{Rots2004}) 
and SWIFT (\citealt{Cusumano2012}) observatories. For NUSTAR the measured correction of 
$5.76 \pm 0.13$ ms was used, which also includes a UTC clock offset (this issue is 
currently under discussion with the NUSTAR help desk). The typical rms error on the TOA
for the three observatories was $34$ $\mu$s, $136$ $\mu$s, and $750$ $\mu$s, respectively.
\begin{table}
\begin{center}
\caption{Pre-glitch reference timing model obtained using $74$ phase residuals  
$\approx 500$ days before CPG2011, at reference epoch MJD $55875.5$. The error in the last digit 
of each number is shown in brackets. \label{tbl1}
}
\begin{tabular}{|l|c|}
\hline
Parameter  & Value \\
\hline
$\nu_0$ (Hz) & $29.706643799(1)$ \\
\hline
$\dot \nu_0$ (Hz s$^{-1}$) & $-370723.9(1) \times 10^{-15}$ \\
\hline
$\ddot \nu_0$ (Hz s$^{-2}$) & $1.1985(5) \times 10^{-20}$ \\
\hline
\hline
\end{tabular}
\end{center}
\end{table}

The $74$ phase residuals  $\approx 500$ days before CPG2011 were fit in TEMPO2
(\citealt{Hobbs2006}) to obtain the pre-glitch reference timing model (see discussion 
below), which is given in Table~\ref{tbl1}; only the last number is not 
statistically consistent with the values derived in section $2$, but it is in the 
same ballpark. The rms of the fit is $327$ $\mu$s. If the data cadence was sufficient 
(e.g., once a day), then the TEMPO2 
phase residuals for the post-CPG2011 TOA would have been consistent with the integral 
of the negative of equation $2$, which is
\begin{equation}
\begin{array}{ll}
\delta \phi(t)  & = \Delta \phi_0 - (1/\nu_0) 
       \left [ \Delta \nu_p t - \Delta \nu_n \tau_1 \left (\exp-(t/\tau_1) - 1 \right ) \right .  \\
              & - \left . \Delta \dot \nu_p \left ( w \tau_2 \left ( 
                   -\tau_2 \left ( \exp-(t/\tau_2) - 1 \right ) - t \right ) + t^2/2
                  \right ) \right ], \\
\end{array}
\end{equation}
\noindent where $\delta \phi$ is measured in seconds. However, the low data cadence in this 
work requires that integral number of periods in time (or cycles in phase) must be added or
subtracted from the TEMPO2 phase residuals in order to match with equation $3$. This was done (if
required) for each  post-glitch residual under the requirements that the modified phase 
should be as close to Eq. $3$ as possible and that the difference between two consecutive 
modified phases should be less than half a cycle (of either sign), the criterion also used 
internally in TEMPO2. For  this, a special plugin was developed in TEMPO2, which plots 
equation $3$ over the data as a guide for inserting the appropriate number of phase cycles. 
This scheme worked for up to $1450$ days after CPG2011, which is sufficient for our purpose.
Beyond $1450$ days the difference between consecutive phase residuals differs by more than 
half a cycle.

This technique is merely a modification of the usual method of using TEMPO2, viz., of using 
the pre-glitch reference timing model on the post-glitch TOAs. If the data cadence was very 
good, e.g., once a day, then one would have immediately obtained the curve describe by 
equation $3$. Given the low data cadence in our data, TEMPO2 has to be aided by manually
inserting the integer number of phase cycles (positive or negative) between consecutive phase
residuals. This is achieved by using equation $3$ as a guide.

\begin{figure}[t]
\centering
\includegraphics[width=8.5cm]{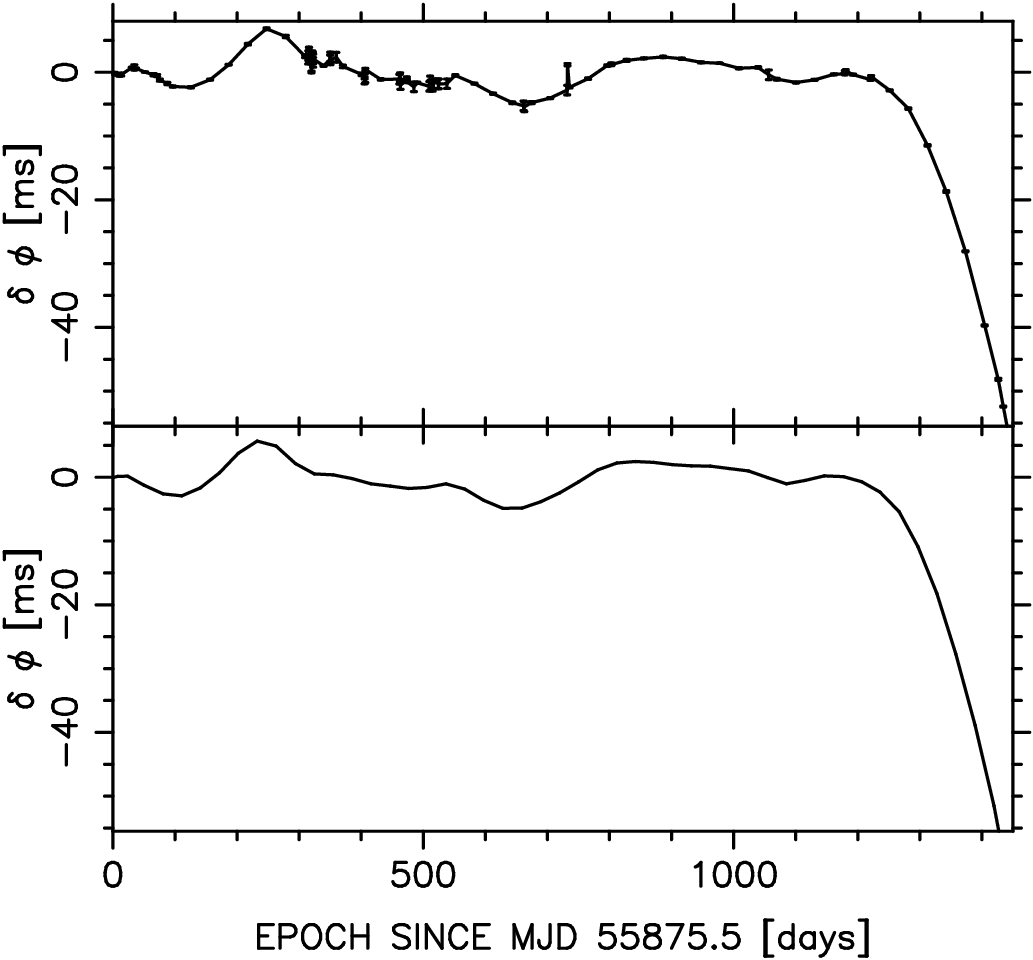}
\caption{
         Top panel: Phase residuals $\delta \phi$ (in milli seconds) between TOA and 
         equation $3$,modified as described in the text. The difference between the 
         reference timing models of sections $4$ and $2$ has been taken into account.
         Bottom Panel: Integration of the negative of the data in the bottom panel of 
         Figure~\ref{fig2}.
        }
\label{fig3}
\end{figure}

The top panel of Figure~\ref{fig3} shows modified phase residuals $\delta \phi$, after 
removing a small linear trend from $0$ to $1200$ days, which implies a small correction of 
$1.77(1) \times 10^{-8}$ Hz in period to the reference timing model. This is most probably
on account of errors on the parameters of the pre-glitch reference timing model, errors on 
the parameters of eq. 3, etc.  It is clear that beyond  epoch  $\approx 1200$ days, the phase 
reduces monotonically, as expected from Figure~\ref{fig2}. 

The reliability of this method is verified in the bottom panel of Figure~\ref{fig3}, 
which shows the integration of $-\delta \nu$ in the bottom panel of Figure~\ref{fig2}, 
using the Trapezoidal rule, and taking the non-uniform spacing of the data epochs into 
account. Even though the two curves are expected to be similar, the actual similarity is
remarkable.

%__________________________________________________________________

%______________________________________________________________

\section{Discussion}

In summary, this work has demonstrated that the Crab Pulsar experienced 
%\LEt{ underwent? "suffered" suggests that it was a bad thing  } 
a speed-up event 
around the end of $2015$ February, that was unlike a glitch or a timing noise behavior. 
It was caused by a persistent increase in $\dot \nu$ of about $\approx 11(1) 
\times 10^{-15}$ Hz s$^{-1}$ after  epoch  $\approx 1200$ days from CPG2011. This  
caused the Crab Pulsar's $\nu$ to increase monotonically by $\approx 47.4 \times 10^{-8}$ 
Hz over $\approx 550$ days.

In Figure~\ref{fig3} the pre-glitch reference timing model was obtained using data 
for the $\approx 500$ days before CPG2011, and not the $800$ days that was used in sections
$1$ and $2$,  because the phase residuals between days $800$ and $500$ 
before CPG2011 showed a significant departure from those between days $500$ and $0$.
It is not possible to state here whether this is on account of poor data cadence or
on account of a genuine sub-event in the Crab Pulsar.

One physical process that can cause a persistent increase in $\dot \nu$ is an increase
in temperature $T$ in the vortex creep regions of the Crab Pulsar (\citealt{Alpar1984}). Vortex 
creep is the mechanism by which  superfluid vortexes move radially outwards steadily, 
thus slowing down the superfluid and speeding up the outer crust, to which the 
radiation that we observe is firmly anchored. Vortexes move radially outwards at speed 
$V_r$, which is a statistical quantity, having both signs in general. However, owing to 
differential rotation between the inner superfluid and the outer crust of a neutron 
star, it is biased towards positive values. Thus its average value $\left < V_r 
\right >$ is greater than $0$, and depends exponentially upon the $T$. A 
change in temperature $\delta T$ gives rise to a change in $\left < V_r \right >$, 
which in turn gives rise to a change in $\dot \nu$ according to the  formula
\begin{equation}
\frac{\delta T}{T} \approx \frac{1}{30}  \frac{\delta \left < V_r \right > }{\left < 
V_r \right >} \approx \frac{1}{30}  \frac{\delta \dot \nu }{\left | \dot \nu \right | };
\end{equation}
\noindent see equations $65$ and $22$ in \cite{Alpar1984}. Strictly, $V_r$ is related 
to $\dot \nu_s$, where $\nu_s$ is the frequency of rotation of the superfluid 
(equation $4$ in \citealt{Alpar1984}); $\nu_s$ is related to the observed $\nu$ through
equation $19$ of \cite{Alpar1984}.

Now, the $\dot \nu$ at  epoch $\approx 1200$ days after CPG2011 is equal to $\dot 
\nu_0 + \Delta \dot \nu_p = -370853(6) \times 10^{-15}$ Hz s$^{-1}$, from 
Figure~\ref{fig1}, while the persistent increase in this quantity is $11(1) \times 
10^{-15}$ Hz s$^{-1}$. Therefore, $\delta \dot \nu / \left | \dot \nu \right | $ is 
$\approx 11 / 370853 \approx 3.0(3) \times 10^{-5}$. Thus, the required change in 
temperature is $\delta T / T \approx 10^{-6}$, which appears to be a very small 
quantity. Then why are speed-up events so rare in the Crab Pulsar?

The first reason is that the required change in temperature in equation $4$ may be 
an underestimate. It is obtained by \cite{Alpar1984} by ignoring the second term in 
their equation $16$, which may modify the dependence of $\delta \dot \nu$ on $\delta 
T$ in equation $4$.

The second reason is given in the para following equation $65$ of \cite{Alpar1984}.
The heat capacity and thermal conductivity of relativistic electrons give a very 
small thermal diffusion timescale of about $1$ sec over $100$ meters 
(\citealt{Flowers1976}). Therefore, any heat creating process may not succeed in raising 
the temperature uniformly and persistently over a sufficiently large creep region, 
due to rapid dissipation of heat to neighboring regions.

What is the cause of the sudden increase in temperature in the regions of vortex creep? 
It could be some fluctuation in the vortex creep process itself, since this process can 
generate significant heat (\citealt{Alpar1984}). This fluctuation could be due to
either magnetic reconnection or relatively slow crustal failure that does not lead 
to a glitch.

This speed-up event has important implications for the next large glitch in the Crab Pulsar.
Glitches are supposed to occur when superfluid vortexes unpin catastrophically, which
occurs when the differential rotation between the pinned internal superfluid and outer 
crust builds up to a critical value (equation $11$ in \citealt{Alpar1984}). Clearly, a 
speed-up event reduces differential rotation;  angular momentum is transferred from 
the faster rotating superfluid to the slower crust, which should work against the 
occurrence of (at least) a large glitch. Therefore, one should logically expect this speed-up event to 
be terminated well before the next large glitch in the Crab Pulsar. On the other hand, perhaps the 
magnitude of the speed-up event is so small that the differential rotation may 
continue to build up to its critical value, but more slowly, so the next large glitch in 
the Crab Pulsar may occur much later than expected. Therefore, whether the speed-up event
persists at the time of the next glitch or  is terminated before the next glitch --
and if terminated then before what duration -- is an important clue to understanding the superfluid 
dynamics of the Crab Pulsar.

When is the next large glitch expected to occur in the Crab Pulsar? By analyzing the epochs 
of occurrence of the $24$ glitches listed in Table $3$ in \cite{Lyne2015}, and 
including the glitch at epoch $\approx 44900$ listed in Table $3$ in \cite{Wong2001}, 
one finds a periodicity of $\approx 2686 \pm 161$ days ($\approx 7.4$ years) for the 
occurrence of glitches in the Crab Pulsar, buried in what otherwise appears to be random
occurrence. In particular, all large glitches, defined by $\Delta \nu / \nu > 30 
\times 10^{-9}$ in Table $3$ in \cite{Lyne2015}, occur at epochs that are at 
intervals of $\approx 2686 \pm 161$ days, or multiples of it, from MJD $42447.26$
($1975$ February), which is the epoch of occurrence of the first recorded large glitch in 
the Crab Pulsar. If one assumes that these statistics are stationary, and that the era of
frequent glitching in the Crab Pulsar is over, as is apparent from the data, and that the
speed-up event has a negligible effect on the statistics, then the next 
large glitch in the Crab Pulsar is expected to occur $\approx 2686$ days after the last 
large glitch, which would imply around MJD $55875.5 + 2686 \approx 58561$, or around 
$2019$ late March, with an rms uncertainty of $\approx 161$ days. Whether this glitch 
will be large or small, and how much later or earlier than $2019$ late March it will  occur, 
will be determined by the effect of the speed-up event on achieving critical differential
rotation.

\begin{acknowledgements}
I thank Sergio Campana for detailed help in enhancing the clarity of and shortening  this paper.
This research made use of data obtained from the High Energy Astrophysics Science Archive 
Research Center Online Service, provided by the NASA-Goddard Space Flight
Center.
\end{acknowledgements}

%-------------------------------------------------------------------

\end{document}